\documentclass[12pt]{article}
\begin{document}
\title{The Dark Energy Paradigm}
\author{B.G. Sidharth\\
International Institute for Applicable Mathematics \& Information Sciences\\
Hyderabad (India) \& Udine (Italy)\\
B.M. Birla Science Centre, Adarsh Nagar, Hyderabad - 500 063
(India)}
\date{}
\maketitle

\begin{abstract}
Though the concept of a dark energy driven accelerating universe was
introduced by the author in 1997, to date dark energy itself, as
described below has remained a paradigm.
\end{abstract}
\section{Introduction}
The author in 1997 proposed a dark energy driven accelerating
universe with a small cosmological constant. In 1998, the
observations of Perlmutter and others on distant type Ia supernovae
confirmed the above scenario - this work was infact the Breakthrough
of the Year 1998 of the American Association for Advancement of
Science's Science Magazine \cite{perl,kirsh,science1}. Subsequently
observations by the Wilkinson Microwave Anisotropy Probe (WMAP) and
the Sloan Digital Sky Survey confirmed the predominance of the new
paradigmatic dark energy - this was the Breakthrough of the Year
2003 \cite{science2}. Such a background energy with negative
pressure (to cause repulsion) is now called Dark Energy. Moreover
the so called Large Number coincidences including the mysterious
Weinberg formula are deduced in the author's theory, rather than
being miraculous coincidences. We examine the concept of dark energy
in this context and indicate how this energy may even be harnessed.\\
We first observe that the concept of a Zero Point Field (ZPF) or
Quantum Vacuum (or Vacuum energy) is an idea whose origin can be
traced back to Max Planck himself. Quantum Field Theory attributes
the ZPF to the virtual Quantum effects of an already present
electromagnetic field \cite{davydov}.\\
In a very intuitive way Faraday could conceive of magnetic effects
in vacuum in connection with his experiments on induction. Based on
this, an aether was used for the propagation of electromagnetic
waves in Maxwell's Theory of electromagnetism, which infact laid the
stage for Special Relativity. This aether was a homogenous,
invariable, non-intrusive, material medium which could be used as an
absolute frame of reference, atleast for certain chosen observers.
However the experiments of Michelson and Morley towards the end of
the nineteenth century, lead to its downfall, and thus was born
Einstein's Special Theory of Relativity in which there is no such
absolute frame of reference.\\
Very shortly thereafter the advent of Quantum Mechanics lead to its
rebirth in a new and unexpected avatar \cite{bgsmyst}. Essentially
there were two new ingredients in what is today called the Quantum
Vacuum. The first was a realization that Classical Physics had
allowed an assumption to slip in unnoticed: In a source or charge
free "vacuum", one solution of Maxwell's Equations of
electromagnetic radiation is no doubt the zero solution. But there
is also a more realistic non zero solution. That is, the
electromagnetic radiation
does not necessarily vanish in empty space.\\
The second ingredient was the mysterious prescription of Quantum
Mechanics, the Heisenberg Uncertainty Principle, according to which
it would be impossible to precisely assign momentum and energy on
the one hand and spacetime location on the other. Clearly the
location of a vacuum with no energy or momentum cannot be specified
in spacetime.\\
This leads to what is called a Zero Point Field. For instance a
Harmonic Oscillator, a swinging pendulum for example, according to
classical ideas has zero energy and momentum in its lowest position.
But the Heisenberg Uncertainty endows it with a fluctuating energy.
This fact was recognized by Einstein himself way back in 1913 who,
contrary to popular belief, retained the concept of aether though
from a different perspective \cite{ra}. It also provides an
understanding of the fluctuating electromagnetic field in vacuum.\\
This mysterious Zero Point Field or Quantum Vacuum energy has since
been experimentally confirmed in effects like the Casimir effect
which demonstrates a force between uncharged parallel plates
separated by a charge free medium, the Lamb shift which demonstrates
a minute oscillation of an electron orbiting the nucleus in an
atom-as if it was being buffetted by the Zero Point Field- the
anomalous Quantum Mechanical gyromagnetic ratio $g = 2$ and so on
\cite{rb}-\cite{rg,mwt}.\\
The Quantum Vacuum is a violent medium in which charged particles
like electrons and positrons are constantly being created and
destroyed, almost instantly, within the limits permitted by the
Heisenberg Uncertainty Principle for the violation of energy
conservation. There are also claims that the virtual photons of the
Quantum Vacuum have been realized as real photons, in an endorsement
of the dynamical Casimir effect (Cf.ref.\cite{wilson}). One might
call the Quantum Vacuum as a new state of matter, a compromise
between something and nothingness. Something which corresponds to
what the Rig Veda described thousands of years
ago: "Neither existence, nor non existence." \\
The Quantum Vacuum can be considered to be the lowest state of any
Quantum field, having zero momentum and zero energy. The properties
of the Quantum Vacuum can under certain conditions be altered, which
was not the case with the erstwhile aether. In modern Particle
Physics, the Quantum Vacuum is responsible for phenomena like Quark
confinement, a property whereby it would be impossible to observe an
independent or free Quark, the spontaneous breaking of symmetry of
the electroweak theory, vacuum polarization wherein charges like
electrons are surrounded by a cloud of other oppositely charged
particles tending to mask the main charge and so on. There could be
regions of vacuum fluctuations comparable to the domain structures
of feromagnets. In a ferromagnet, all elementary electron-magnets
are aligned with their spins in a certain direction. However there
could be special regions wherein the spins are aligned
differently.\\
Such a Quantum Vacuum can be a source of cosmic repulsion, as
pointed by Zeldovich and others \cite{rh,ri}. However a difficulty
in this approach has been that the value of the cosmological
constant turns out to be huge, far beyond what is observed. This has
been called the cosmological constant problem \cite{rj}. If true, the universe
would have exploded into nothing, shortly after its birth.\\
There is another approach, sometimes called Stochastic
Electrodynamics which treats the ZPF as primary and attributes to it
Quantum Mechanical effects \cite{santos,depena}. It may be
re-emphasized that the ZPF results in the well known experimentally
verified
Casimir effect \cite{mostepanenko,lamor}.\\
We would first like to observe that the energy of the fluctuations
in the background electromagnetic field could lead to the formation
of elementary particles. Indeed this was Einstein's belief. As he
observed as early as 1920 itself  \cite{r22}, "... according to our
present conceptions, the elementary particles are... but
condensations of the electromagnetic field."\\
In the words of Wilzeck, \cite{r20}, "Einstein was not satisfied
with the dualism. He wanted to regard the fields, or ethers, as
primary. In his later work, he tried to find a unified field theory,
in which electrons (and of course protons, and all other particles)
would emerge as solutions in which energy was especially
concentrated, perhaps as singularities. But his efforts in this
direction did not lead to any tangible success."
\section{The Quantum Vacuum}
Let us consider, following Wheeler \cite{mwt} a harmonic oscillator
in its ground state. The probability amplitude is
$$\phi (x) = \left(\frac{m \omega}{\pi \hbar}\right)^{1/4}
e^{-(m\omega / 2 \hbar)x^2}$$
for displacement by the distance
from its position of classical equilibrium. So the oscillator
fluctuates over an interval
$$\Delta x \sim (\hbar / m \omega)^{1/2}$$
The electromagnetic field - or for that matter, any collection of
bosons - is an infinite collection of independent oscillators, with
amplitudes $X_1, X_2$ etc. The probability for the various
oscillators to have amplitudes $X_1 , X_2$ and so on is the product
of individual oscillator amplitudes:
$$\phi (X_1,X_2 \cdot ) = exp \left[ - \left(X^2_1 + X^2_2 + \cdot
\right)\right]$$ wherein there would be a suitable normalization
factor. This expression gives the probability amplitude  $\phi$  for
a configuration $B(x,y,z)$  of the magnetic field that is described
by the Fourier coefficients $X_1,X_2,\cdots$ or directly in terms of
the magnetic field configuration itself we have
$$\phi (B(x,y,z)) = P exp \left(- \int \int \frac{B(x_1) \cdot
B(x_2)}{16 \pi^3 \hbar cr^2_{12}}\right)$$ $P$  being a
normalization factor. Let us consider a configuration where the
magnetic field is everywhere zero except in a region of dimension
$l$, where it is of the order of $\sim \Delta B$. The probability
amplitude for this configuration would be proportional to
$$exp [ - (\Delta B)^2 l^4 / \hbar c)$$
So the energy of fluctuation in a region of length $l$ is given by
finally \cite{mwt,r24,r25}
\begin{equation}
B^2 \sim \frac{\hbar c}{l^4}\label{e1}
\end{equation}
We next argue that $l$, the mean length of fluctuations,  will be
the Compton length.  We note that as is well known, a background ZPF
of the kind we have been considering can explain the Quantum
Mechanical spin half as also the anomalous $g = 2$ factor for an
otherwise purely classical electron \cite{sachi,boyer,taylor}. The
key point here is (Cf.ref. \cite{sachi}) that the classical angular
momentum $\vec{r} \times m\vec{v}$ does not satisfy the Quantum
Mechanical commutation rule for the angular momentum $\vec{J}$.
However when we introduce the background Zero Point Field, the
momentum now becomes
\begin{equation}
\vec{J} = \vec{r} \times m = \vec{v} + (e / 2c) \vec{r} \times
(\vec{B} \times \vec{r} + (e / c) \vec{r} \times
\vec{A}^0,\label{e2}
\end{equation}
where $\vec{A}^0$ is the vector potential associated with the ZPF--
for example if the electric part of the ZPF is $\vec{E}^0$, this is
usually considered to be a Gaussian random process and $\vec{A}^0$
is related to $\vec{E}^0$  by the usual Maxwell equation. $\vec{B}$
is an external magnetic field introduced merely for convenience, and
which can be made vanishingly small.\\
It can be shown that $\vec{J}$ in (\ref{e2}) satisfies the Quantum
Mechanical commutation relation for $\vec{J} \times \vec{J}$. At the
same time we can deduce from (\ref{e2})
\begin{equation}
\langle J_z \rangle = \frac{1}{2} \hbar \omega_0 / | \omega_0
|\label{e3}
\end{equation}
Relation (\ref{e3}) gives the correct Quantum Mechanical results
referred to above. From (\ref{e2}) we can extend the arguments and
also deduce that
\begin{equation}
l = \langle r^2 \rangle^{\frac{1}{2}} =
\left(\frac{\hbar}{mc}\right)\label{e4}
\end{equation}
(\ref{e4}) shows that the mean dimension of the region in which the
fluctuation contributes is of the order of the Compton wavelength of
the electron. By relativistic covariance (Cf.ref.\cite{sachi}), the
corresponding time scale is at the Compton scale.\\
In (\ref{e1}) above if $l$ is taken to be the Compton wavelength of
a typical elementary particle, then we recover its energy $mc^2$, as
can be easily verified. As mentioned Einstein himself had believed
that the electron was a result of such condensation from the
background electromagnetic field (Cf.\cite{r26,ri} for details).\\
There could be a cosmological signal of dark energy. As it is the ZPF, we recall that the ZPF causes the Lamb Shift (as well as, via Zitterbewegung, the Darwin term). This, in the Hydrogen atom is $\sim 1000 MHz$. Thus we could expect that the ZPF would leave isotropic radio waves of wavelength in the metre region, not tied to any specific radio source in the sky.
\section{Cosmology}
We now very briefly indicate the cosmology referred to in the
introduction (Cf.ref.\cite{bgskluwer,csf}). But before that we
summarize the new cosmos that has emerged since 1997: It is
essentially flat with its energy constant estimated as, around $4
\%$ ordinary matters some $25\%$ of as yet undetected dark matterm,
while the rest is homogenously spread out dark energy. Returning to
our model, elementary particles are created from the ZPF as above.
If there are $N$ elementary particles, then fluctuationally a nett
$\sqrt{N}$ particles are created within the Compton time $\tau$ (see
ref.\cite{bgskluwer} for details), so that
\begin{equation}
\frac{dN}{dt} = \frac{\sqrt{N}}{\tau}\label{e5}
\end{equation}
We also use the well known facts that
\begin{equation}
M = Nm\label{e6}
\end{equation}
and
\begin{equation}
R = GM / c^2\label{e7}
\end{equation}
In (\ref{e6}), $M$ is the mass of the Universe, $m$ the mass of a
typical elementary particle like the pion, $N \sim 10^{80}$ the
number of elementary particles in the Universe and $R$ its radius.
Differentiation of (\ref{e7}) and use of (\ref{e6}) and (\ref{e5})
then leads to a host of consistent relations,
\begin{equation}
v = \dot{R} = HR, \quad H = \frac{c}{l} \cdot
\frac{1}{\sqrt{N}},\label{e8}
\end{equation}
\begin{equation}
G\rho_{vac} = \Lambda < 0 (H^2), R = \sqrt{N} l, T = \sqrt{N} \tau ,
\rho_{vac} = \rho / \sqrt{N}\label{e9}
\end{equation}
\begin{equation}
m = \left(\frac{H\hbar^2}{Gc}\right)^{1/3} , \frac{e^2}{Gm^2}
\approx \frac{1}{\sqrt{N}}\label{e10}
\end{equation}
and so on.\\
In (\ref{e8}) above, $H$ is the Hubble constant, $l$ the pion
Compton length, while in (\ref{e9}) $\rho$ the average density,
$\Lambda$ the cosmological constant and $\rho_{vac}$ the vacuum
density. The second relation of (\ref{e9}) is the empirically known
so called Eddington formula. The first and second relations of
(\ref{e10}) are respectively, the Weinberg formula and the well
known (but otherwise ad hoc) electromagnetism - gravitational
coupling
constant.\\
It may also be mentioned that all this can be interpreted elegantly
in terms of underlying Planck oscillators in the Quantum Vacuum
(Cf.refs.\cite{fpl2,fpl3}).\\
Finally, it may be mentioned that (\ref{e10}) shows that both
$\Lambda$ and $H \to 0$ as $N \to \infty$, as indeed is the current
belief.
\section{Harnessing the ZPF?}
Two of the earliest realizations of the Vacuum energy as mentioned
were in the form of the Lamb shift and the Casimir effect.\\
In the case of the Lamb shift, as is well known, the motion of an
orbiting electron gets affected by the background ZPF. Effectively
there is an additional field, over and above that of the nucleus.
This additional potential, as is well known is given by \cite{bd}
$$\Delta V (\vec{r}) = \frac{1}{2} \langle (\Delta r)^2 \rangle
\nabla^2 V (\vec{r})$$ The additional energy
$$\Delta E = \langle \Delta V (\vec{r})\rangle$$
contributes to the observed Lamb shift which is $\sim 1000 mc /
sec$. The essential idea of the Casimir effect is that the
interaction between the ZPF and matter leads to macroscopic
consequences. For example if we consider two parallel metallic
plates in a conducting box, then we should have a Casimir force
given by \cite{r3}
$$F = \frac{- \pi^2 \hbar cA}{240 l^4}$$
where $A$ is the area of the plates and $l$ is the distance between
them. More generally, the Casimir force is a result of the
boundedness or deviation from a Euclidean topology in the Quantum
Vaccuum. These Casimir forces have been experimentally demonstrated
\cite{r4,r5,r6,r7}.\\
Let us return to equation (\ref{e1}). The ZPF fluctuations typically
take place within the time $\tau$, a typical elementary particle
Compton time as suggested by (\ref{e4}). This begs the question
whether such ubiquotous fields could be tapped for terrestrial
applications or otherwise. We now invoke the well known result from
macroscopic physics that the current in a coil is given by
\begin{equation}
\imath = \frac{nBA}{r\Delta t}\label{e11}
\end{equation}
where $n$ is the number of turns of the coil, $A$ is its area and
$r$ the resistance.\\
Introducing (\ref{e1}) into (\ref{e11}) we deduce that a coil in the
ZPF would have a fluctuating electric current given by
\begin{equation}
\imath = \frac{nAe}{r l^2 r}\label{e12}
\end{equation}
Of course, this would be a small effect. But in principle it should
be possible to harness the current (\ref{e12}) using advanced
technologies, possibly superconducting coils to minimize $r$.
\section{Discussion and a Model For the Cosmological Constant}
Although the concept of dark energy, is now taken for granted, its
exact characterization is still a mystery. Very broadly there are
two approaches. One is the cosmological constant approach we saw
above. The other is to identify dark energy with a scalar field, for
instance quintessence. Such a field can also be associated with a
particle, fundamental or composite. Tachyonic fields have also been considered.\\
For example we could consider an interaction of dark energy with a
fermionic field, contained in dark matter, these fermions being
neutrinos \cite{dela,corasaniti}. Attempts have been made to
formulate an equation of state for a dark energy fluid \cite{Shin}.
Questions have also been asked whether we have dissipated cosmology
or conservative cosmology as a result \cite{marek}, while a
generalized second law has also been studied \cite{German}. The
coincidence problem is also being studied viz., why the energy
density of dark energy is roughly of the same order as a
cosmological critical density \cite{Straumann,Urbano,Wen}.\\
The question may also be asked, what of dark matter, which has
defied observation even after 75 years? Indeed the nature of dark
matter is yet unresolved, assuming that it exists. In the author's
cosmology discussed above we have a gravitational constant that
depends inversely on time, as can be seen, for example from equation
(\ref{e10}). The author has argued over the years that such a time
varying gravitational constant can explain the observational
anomalies which are sought to be explained by dark matter, for
example the flattening of the galactic rotation curves and so on
(Cf. ref. \cite{tduniv} and several references therein). Indeed we
will show that equation (\ref{e10}) ultimately leads to a uniform
cosmic acceleration of the rough order of $10^{-7} cm / sec^2$. In
this sense this approach is a substitute for the ad hoc modified
Newtonian gravity approach.\\
In earlier communications it was shown, on the basis of the cold
cosmic neutrino background, that we can consistently get the
neutrino mass and other neutrino parameters
\cite{bgsfpl,hayakawa,century}. The neutrino mass thus obtained is
in agreement with the value obtained from the SuperKamiokande
experiments-- and infact predicted these results \cite{bgsmass}. One
way of seeing this is to consider the cold Fermi degenerate gas
\cite{huang}. We have
\begin{equation}
p^2_F = \hbar^3 (N/V)\label{ex}
\end{equation}
Feeding in the known neutrino parameters, viz., \cite{ruffini} $N
\sim 10^{90}$ we get from the above, the correct neutrino mass $\sim
10^{-3} eV$ and the background temperature $T \sim 1^\circ K$.  More
recently there has been hope that
neutrinos can also exhibit the ripples of the early Big Bang and in fact, Trotta and Melchiorri claim to have done so \cite{trotta}.\\
\indent It may be mentioned that there is growing evidence for the
cosmic background neutrinos \cite{weiler}. The GZK photo pion
process seems to be the contributing factor.\\
With this background we try to extract the cosmological constant
from the Fermi energy of the neutrinos background. We have
\begin{equation}
\mbox{Fermi \, Energy}\, = \frac{N^{5/3} \hbar^2}{m_\nu R^2} = M
\Lambda R^2\label{XA}
\end{equation}
From (\ref{XA}) we get,
\begin{equation}
\Lambda = 10^{-37}\label{XB}
\end{equation}
which gives the correct order of the cosmological constant.

\end{document}